\documentclass[12pt]{article}

\usepackage[dvips]{graphicx}
\usepackage{amssymb,amsfonts,amsmath,latexsym}
\usepackage{epsfig, graphicx}

\usepackage{amstext}
\usepackage{epsfig, graphicx}
\usepackage{epsfig}
\usepackage{latexsym}
\usepackage{amstext}
\usepackage{amssymb}
\usepackage{graphics}
\usepackage{color}

\newcommand{\beq}{\begin{eqnarray}}
\newcommand{\beqq}{\begin{eqnarray*}}
\newcommand{\eeq}{\end{eqnarray}}
\newcommand{\eeqq}{\end{eqnarray*}}

\definecolor{red}{rgb}{1,0,0}

\begin{document}
\title{Potential wells for AMPA receptors organized in ring nanodomains}
\author{ N. Hoze$^{1}$  D. Holcman$^{1}$ \footnote{$^1$Group of Computational Biology and Applied Mathematics, Institute of  Biology, Ecole Normale Sup\'erieure, 46 rue d'Ulm 75005 Paris, France}}
\maketitle

\begin{abstract} 
By combining high-density super-resolution imaging with a novel stochastic analysis, we report here a peculiar nanostructure organization revealed by the density function of individual AMPA receptors moving on the surface of cultured hippocampal dendrites.  High density regions of hundreds of nanometers for the trajectories are associated with local molecular assembly generated by direct molecular interactions due to physical potential wells. We found here that for some of these regions, the potential wells are organized in ring structures. We could find up to 3 wells in a single ring. Inside a ring receptors move in a small band the width of which is of hundreds of nanometers.  In addition, rings are transient structures and can be observed for tens of minutes. Potential wells located in a ring are also transient and the position of their peaks can shift with time. We conclude that these rings can trap receptors in a unique geometrical structure contributing to shape receptor trafficking, a process that sustains synaptic transmission and plasticity.
\end{abstract}

{\bf Keywords: AMPAR trafficking | Super-resolution data | Stochastic analysis of trajectories | Single Particle Tracking | Lateral Diffusion | Molecular Interactions}
\section*{Introduction}\label{ sec:intr}
Receptor trafficking has been identified as a key feature of synaptic transmission and plasticity \cite{Nicoll0,Nicoll1,Bredt1,Tomita,Schnell2002,Malinow2,Malinow3}. Yet, the mode of trafficking remains unclear: after receptors are inserted in the plasma membrane of neuron, classical single particle tracking revealed that the motion can either be free or confined Brownian motion \cite{Choquet0,Choquet-triller}. Recently,  super-resolution light optical microscopy techniques for \textit{in vivo} data \cite{Choquet2,HozePNAS} have allowed monitoring a large number of molecular trajectories at the single molecular level and at nanometer resolution. Using a novel stochastic analysis approach \cite{HozePNAS}, we have identified that high confined density regions are generated by large potential wells (100s of nanometers) that sequester receptors. In addition, fluctuation in the apparent diffusion coefficient reflects the change in the local density of obstacles \cite{Hoze2}.

Classically, cell membranes are organized in local microdomains \cite{Sheetz,KusumiReview} characterized by morphological and functional specificities. In neurons prominent microdomains include dendritic spines and synapses, which play a major role in neuronal communication. In this report, we analyze  AMPAR, that are key in mediating transmission in excitatory glutamatergic transmission, are not only relocating between synaptic and extrasynaptic sites due to lateral diffusion \cite{Bredt1,Choquet0,Choquet1}, but can be trapped in transient ring structures that contain several potential wells.

We identified three regions (Fig. 1) where trajectories form rings (Fig. 2A-B): specifically, a first feature that defines a ring is that receptor trajectories are constraint into an annulus type geometry (Fig. 2C) located on the surface of the membrane. However, rings are not necessarily localized within dendritic spines. To further characterize a ring, we use the stochastic analysis developed in \cite{HozePNAS}, to compute the vector field for the drift, which is surprisingly restricted to an annulus (Fig. 2B, Ring 2 and 3): the size of the inner radius is of the order of 500 nm while the external radius is around $1 \mu m$. In addition, inside a ring, we found several attracting potential wells of various sizes that co-localized with regions of high density (Fig. 3A).  From the density distribution function along the ring  (Fig. 3), we identified three wells. While the AMPARs diffusion coefficient in these wells is around $0.6 $ $\mu$m$^2$/s (Fig. 3B), the wells have an interaction range of about 500 nm (Fig. 3C), and are associated with an energy of 3.6 $kT$, 1.8 $kT$ and 3 $kT$ respectively.

Using time lapse imaging (Fig. 2), we could see that a ring can be stable over a period of 30 minutes (Ring 1), while ring 2 was only stable for 15 minutes. Interestingly, we could see a transient ring which creates a transient structure interconnecting two-ring like structures at time 45 minutes before it disappears at 60 minutes. Ring 3 could be detected transiently after 15 minutes, containing multiple wells.

Finally, focusing on Ring 1, by plotting the density function of trajectories, we could see over time  how a potential well emerges, is destroyed  as well as its stability (Fig. 4A). While the diffusion coefficient remains constant over time (Fig. 4B), the energy of the main well (fig. 4C) changes across the time lapse experiments as follow:  at time 30 min it is E = 6.6 kT (score = 0.25 and a depth of A=2.0), at the intermediate time 45 min, the energy is E= 7.8 kT (score = 0.20, depth A=2.3) and finally at time 60 minutes E = 5.2 kT (score = 0.13, depth A=1.6). The weak score confirms the likelihood of the well \cite{HozePNAS}. Interestingly, the energy level of the well is neither weak and strong \cite{HozePNAS}, but remains stable overtime.
\subsection*{Conclusion}
To conclude,  potential wells that reflect the interaction of AMPA receptor with molecular partner are not only appearing isolated or at synapses \cite{HozePNAS}, but can also appear in ring structures.  Although these rings can be due to latex bead, the organization of potential wells around in such vicinity suggest that membrane curvature could be a key necessary component to shape the strength of the wells. These rings can trap receptors in 100 nanometers structures. Both potential wells and rings are transient, but stable over periods of minutes.

Classical physical modeling and statistical analysis of single receptor motion \cite{KusumiReview,Saxton-95,Saxton-97,Saxton-08b} assumes that the main driving force is free or confined Brownian motion. This is based on Langevin's description of a pointwise stochastic object description. However because gradient forces, such as electrostatic forces, which are the main sources of chemical interactions cannot generate close trajectories in the determinist limit of Langevin's approximation, maintaining receptor in ring cannot be due to electrostatic forces alone. Thus we propose several hypothesis for this ring structure: one possibility is that receptor dynamics description cannot be reduced to a single point and rather we have now to account for the complex structure of the protein that can generate interactions at specific protein group, which is decoupled from the center of mass. For example the C-terminus tail can interact with some supra-structure generated by scaffolding molecules. Another possibility is that rings are due to transient geometrical structures on the membrane, which traps receptors. We could imagine that this is the case near an endosomal compartment during vesicular exocytosis and endocytosis \cite{Ehlers}.

\section*{Comment}
The data analyzed in this report have been previously published in \cite{HozePNAS} and were generated by D. Nair, J.B. Sibarita, E. Hosy and D. Choquet. We thank them for providing us access to these data.



\begin{figure*}[ht!]
\begin{center}
\includegraphics[scale=0.65]{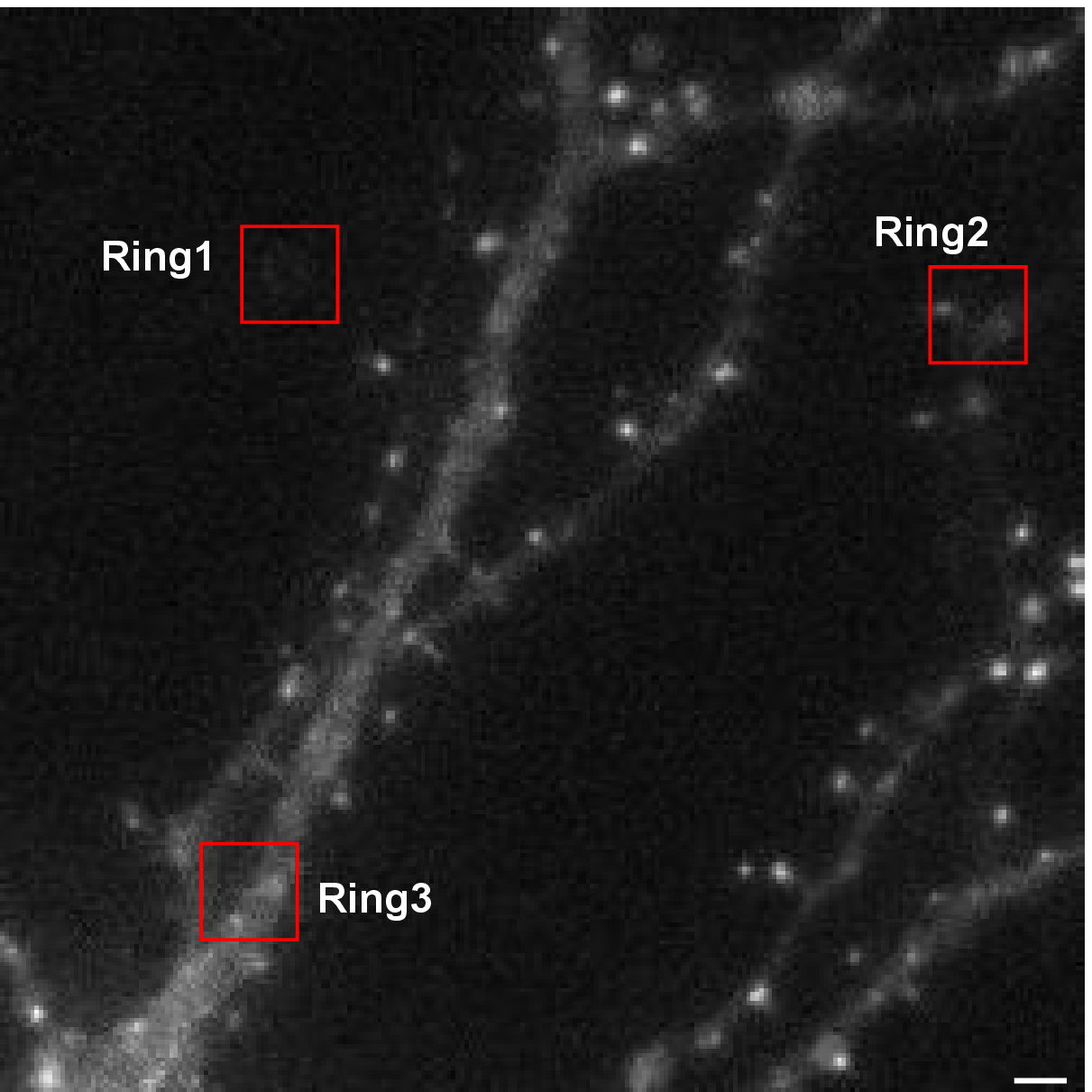}
\caption{ Cultured of hippocampal neurons, where synapse are identified with the marker homer (white spots). Three square areas are highlighted where we identified ring structures. Image was acquired before sptPALM experiments.  \textbf{Scale bar 2 $\mu$m} }
\label{fig1}
\end{center}
\end{figure*}

\begin{figure}[ht!]
\begin{center}
\includegraphics[scale=0.7]{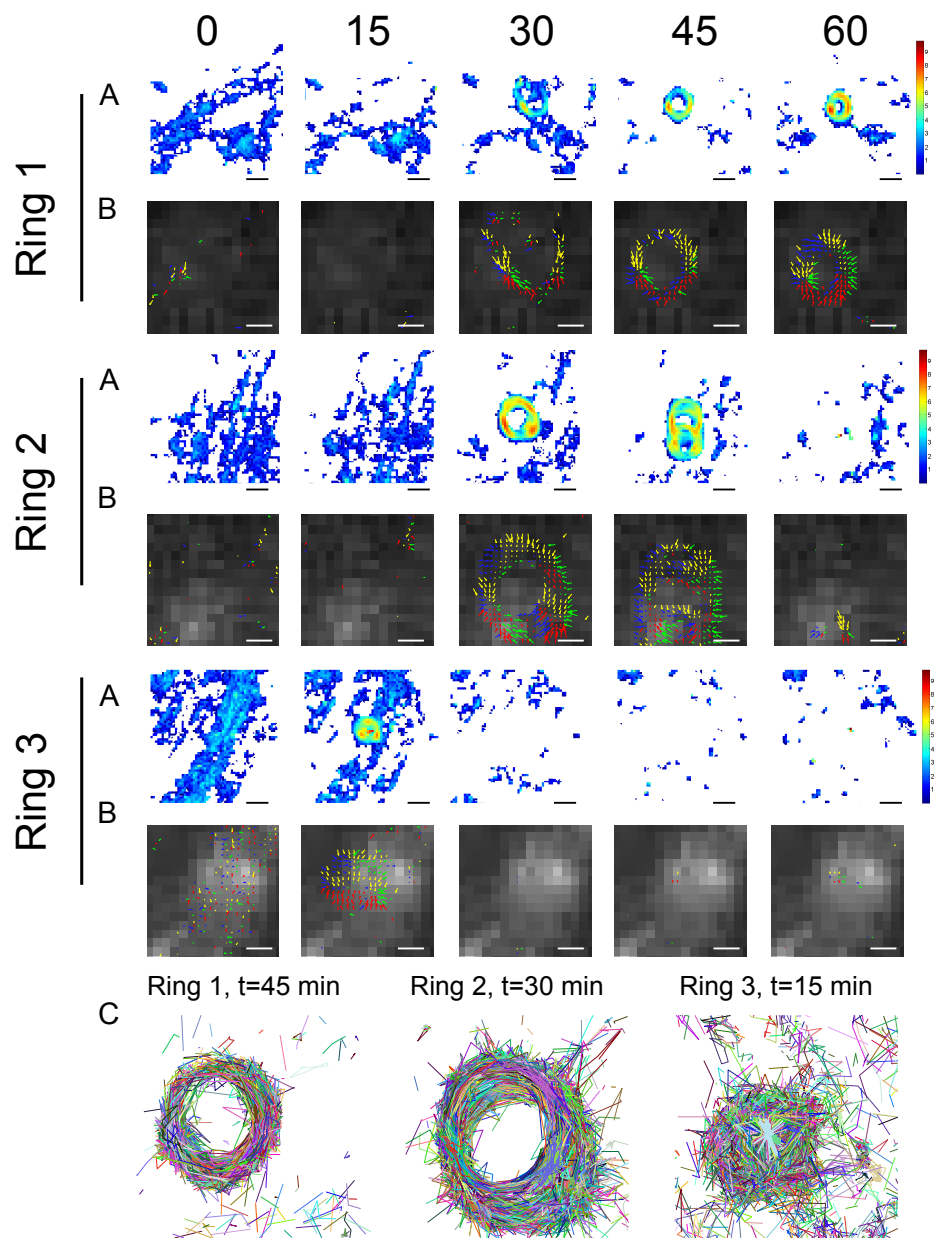}
\caption{ {{\bf Time lapse images of several wells.}} {Ring 1 Scale bar 500 nm.
Trajectories were acquired during 1 minute every fifteen minutes.
(A) Distribution of the trajectory density in the ring regions defined in Figure 1. Scale bar 1 $\mu$m.
(B) Velocity field in the ring regions. Scale bar 500 nm.
(C) Ensemble of the trajectories acquired at the rings during one minute. Scale bar 500 nm.} }
\label{figtimelapse}
\end{center}
\end{figure}

\begin{figure}[ht!]
\begin{center}
\includegraphics[scale=0.65]{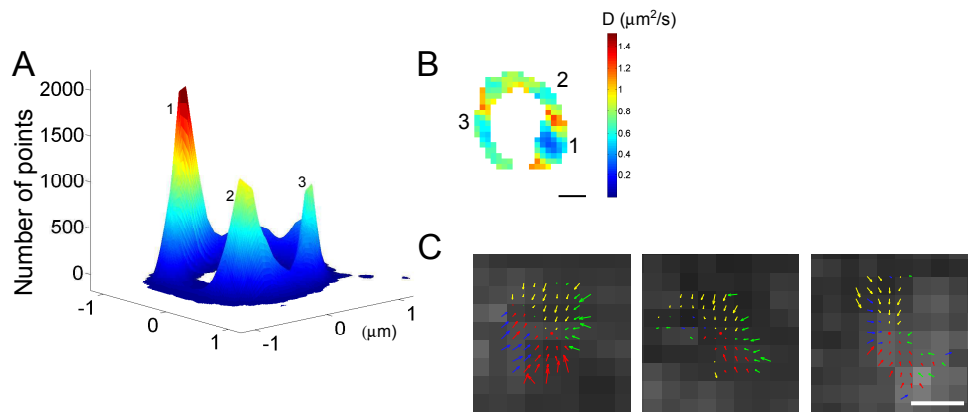}
\caption{ {\bf Features at Ring 2.} The analysis was performed on the dataset taken during 1 minute after 30 minutes of experiments. (A) Distribution of density of trajectories on Ring 2. Three high density regions are identified, associated with three different potential wells.
(B) Map of the diffusion coefficient. Scale bar 1 $\mu$m. (C) Velocity map at the potential wells. Scale bar 500 nm. }
\label{stargazin}
\end{center}
\end{figure}

\begin{figure}[ht!]
\begin{center}
\includegraphics[scale=0.75]{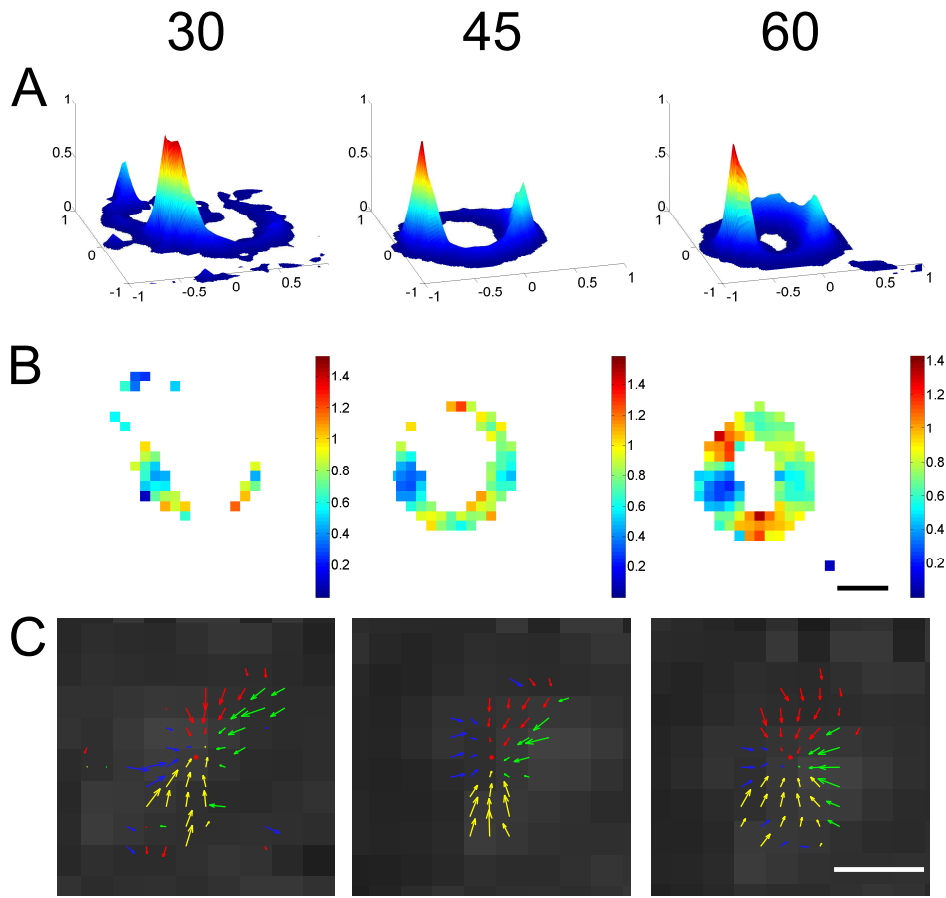}
\caption{ {\bf Time Lapse analysis of potential wells in Ring 1.} The dataset is taken during 1 minute after 30, 45 and 60 minutes of experiments. (A) Distribution of density of trajectories at Ring 1. One potential well is sustained in time, although the position of the peak is slightly shifted between 30 and 45 minutes. Additionally a potential well disappears between 30 and 45 minutes, while another one appears at 45 minutes.
(B) Map of the diffusion coefficient. (C) Velocity map at the potential well. Scale bars 500 nm. }
\label{fig4}
\end{center}
\end{figure}

\end{document}